\documentstyle[aps,prc,preprint]{revtex}

\begin{document}
\def\btau{\vec{\tau}}
\def\bphi{\vec{\phi}}
\draft
\preprint{FTUV-99-50,IFIC-99-52}
\title{Scalar Isoscalar pion pairs in nuclei
and the $A(\pi,\pi\pi)X$ reaction} 
\author{M.J. Vicente Vacas and E. Oset}
\address{Departamento de F\'{\i}sica Te\'{o}rica and IFIC, Centro Mixto 
Universidad de Valencia-CSIC,\\ 46100 Burjassot, Valencia, Spain}
\date{\today}
\maketitle

\begin{abstract}
The reaction $A(\pi,\pi\pi)X$ has been studied at low energies, paying 
particular attention to the interaction of the two final pions in the
scalar isoscalar (I=J=0) channel.
We have developed a microscopic model for the pion production, and then 
implemented the two pion final state interaction by using the results of
a non-perturbative unitary coupled-channels method based in  the standard 
chiral Lagrangians.
The resulting model, describes well the reaction on the nucleon for all 
different isospin channels. 
 Finally, we have considered the reaction in nuclei. 
Our calculation takes into account Fermi motion, Pauli blocking,
pion absorption, and also the strong modification of the $\pi\pi$  
interaction in the nuclear medium.
\end{abstract}


\section{Introduction}

The reaction $A(\pi,\pi\pi)X$ has attracted much interest lately, as a means 
to investigate the properties of correlated $I=J=0$ pion pairs, 
the "$\sigma$" meson, in nuclear matter\cite{Rap:99,Hat:xx}. This has been 
partly motivated by
recent experimental results, which show an $A$ dependent large enhancement 
of the cross section at low invariant masses of the dipion system
\cite{Bon:96,Bon:98}.
 The enhancement occurs for the $A(\pi^-,\pi^+\pi^-)X$ process 
($I_{\pi\pi}=0$ and 2 at low energies), but is conspicuously absent for 
the $A(\pi^+,\pi^+\pi^+)X$ case ($I_{\pi\pi}=2$). Thus, it 
happens only when $I_{\pi\pi}=0$ pion pairs are produced, and could be 
related to the medium dependence of the $\pi\pi$ interaction in the
$\sigma$ channel.

 Our theoretical understanding of the $\pi\pi$ scattering in vacuum has 
improved considerably in the last few years. 
Close to threshold, the $\pi\pi$ interaction is well described  by Chiral 
Perturbation Theory ($\chi$PT). At higher energies, where unitarity matters 
and we are out of the $\chi$PT range of validity, some very successful non 
perturbative models have been developed\cite{Jul:90,Oll:97}. They
are able to give a good description of the $\pi\pi$ scattering
in the $\sigma$ channel up to energies above 1 GeV.
 
Whereas in vacuum the low energy $\pi\pi$ interaction in the $I=J=0$ channel, 
although attractive, is too weak to  admit a bound state, it was soon
realized that the nuclear medium attraction of the pions could lead to the
accumulation of  some strength close to the two pion threshold, or even 
to the appearance of a new bound state\cite{Sch:88}.
A similar conclusion has been reached in Ref.\cite{Hat:xx}, where an enhancement
of the the spectral function in the $\sigma$ channel, just above the 
$2\pi$ threshold, has been predicted as  a consequence of a possible partial
restoration of chiral symmetry in the nuclear medium.
Actually, in more detailed calculations, based on the $\pi\pi$ interaction 
model of ref. \cite{Jul:90} it has been found that spontaneous s-wave
pion pair condensation could appear at densities as low as $\rho_0$
\cite{Cha:91,Mul:92,Aou:93}.
Further works along the same line, but using  the chirally improved
J\"ulich model for the meson-meson interaction, have shown that instabilities 
are pushed up in density when chiral constraints are imposed
\cite{Rap:96,Aou:95}. Nonetheless, some accumulation of strength close to
the two pion threshold was found.
Similar results have been obtained in ref. \cite{Chi:97}. The latter work is
based in the model for  $\pi\pi$ interaction of ref. \cite{Oll:97},
which generates the scattering  amplitude using the chiral lagrangians, and 
then unitarizes it by the inverse amplitude method. 

The first experimental signals showing a small enhancement of the $\pi\pi$
spectral function in the $A(\pi,2\pi)X$ reaction were found in
ref. \cite{Cam:93}, although the experiment didn't measure below 
$M_{\pi\pi}=300$ MeV, and hence, missed the most interesting region.
The advent of CHAOS improved considerably the experimental possibilities,
and it has provided us with a wealth of good quality data for several
charge channels and nuclei. The results show the emergence of a very clear 
peak in the $M_{\pi\pi}$ spectral function close to threshold, when
pions in the scalar isoscalar channels are present in the final state 
\cite{Bon:96,Bon:98,Bon:99}. This peak, absent for the reaction on deuterium
already appears on light nuclei, and grows as a function of the atomic number.
In contrast, only  minor changes are observed when the cross section
of the $(\pi^+,2\pi^+)$ reaction in deuterium is compared to the same process in
heavier nuclei.
Certainly, the different behaviour of the two processes cannot be caused by  
trivial nuclear effects, like Fermi motion, Pauli blocking,
or pion absorption, which are practically identical for both cases.

In ref. \cite{Rap:99}, a calculation of the $A(\pi,2\pi)X$ reaction 
has been presented in which the $\pi\pi$ FSI was taken into account 
using the approach of ref. \cite{Rap:96}. The emphasis being placed on the
analysis of FSI, the elementary mechanism of 
pion production on a single nucleon ($\pi N \to \pi\pi N$) was described 
in a simplified manner, taking only the most important pieces of the
amplitude. Also, some simplifications were done on the treatment of nuclear 
effects, like Fermi motion, or pion absorption. 
The results are very stimulating, and show some enhancement on the
$M_{\pi\pi}$ distribution close to threshold, which agrees well with
the experimental data. 

Our aim in this paper is to do a more detailed study of the reaction, 
including a realistic $\pi N \to \pi\pi N$ amplitude, and incorporating some 
nuclear effects omitted in the previous work, and which could affect its 
conclusions. 

 In sec.\ \ref{sec:chiang} we give a brief outline of the model for 
$\pi\pi$ scattering in the scalar isoscalar channel both in vacuum
\cite{Oll:97} and in nuclear matter \cite{Chi:97}. Then, in sec.\ \ref{sec:pin}
we develop a model for the $\pi N \to \pi\pi N$ reaction. Finally, in sec. 
\ref{sec:nuc} we discuss the process in nuclei. 

\section{Basic theory of the $\pi\pi$ scattering in the scalar isoscalar 
channel}
\label{sec:chiang}
 
\subsection{$\pi\pi$ scattering in  vacuum}

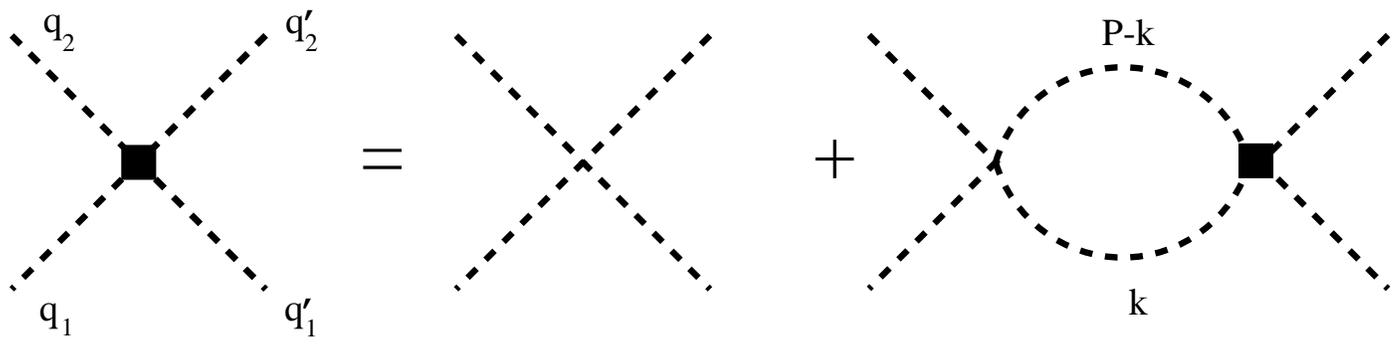
\begin{figure}
\caption{Diagrammatic representation of the Bethe-Salpeter equation.}
\label{fig:LS} 
\end{figure}

The $\pi\pi$ scattering amplitude is calculated solving the coupled channels
Bethe-Salpeter (BS) equation
\begin{equation}
\label{eq:LS}
T_{\pi\pi} = {\mathcal{V}}_{\pi\pi} +{\mathcal{V}}_{\pi l} 
{\mathcal{G}}_{ll} T_{l \pi},
\end{equation}
where the subindex $\pi$ corresponds to the state $|\pi\pi ,I = 0>$
and the subindex $l$ accounts for  the $|\pi\pi ,I = 0>$
and the $| K \bar{K}, I = 0>$ states. The potentials ${\mathcal{V}}_{ij}$ are 
obtained from 
the lowest order chiral Lagrangians\cite{Gas:85}. Explicit expressions can be
found in Ref.\ \cite{Chi:97}.  The 
${\mathcal{VG}}T$ term of eq.\ \ref{eq:LS} stands for 
\begin{equation}
 {\mathcal{VG}}T = \int \frac{d^4 k}{(2 \pi)^4} 
{\mathcal{V}} (q_1, q_2, k){ \mathcal{G} }(P,k) T (k; q'_1, q'_2),
\end{equation}
where $q,q',P$, and $k$ are the momenta of the mesons as defined in 
fig.\ \ref{fig:LS}. A cutoff ($q_{max} = 1 $ GeV) is used in the evaluation of
the integral. $ {\mathcal{G}}(P,k)$ is the product of the two meson propagators  in
the loop,
\begin{equation}
 {\mathcal{G}}_{lj}(P,k) = i \frac{1}{k^2 - m^2_{l} + i \epsilon} \quad 
\frac{1}{(P - k)^2 - m^2_{j} + i \epsilon}.
\end{equation}
The particular ${\mathcal{V}}_{ij}$ off-shell dependence leads to the simplification
of the integral BS equation which can be transformed into the purely 
algebraic expression
\begin{equation}
T_{ik} = V_{ik} + V_{ij} G_{jj} T_{jk}\, ,
\end{equation}
where $V_{il}$ is now the on-shell part of the potentials (when the momenta are
taken such that $p_i^2=m_i^2$), and
\begin{equation}
G_{jj} = i \int \frac{d^4 k}{(2 \pi)^4} \frac{1}{k^2 - m_{1j}^2 + i \epsilon}
\quad \frac{1}{(P - k)^2 - m_{2j}^2 + i\epsilon}\, .
\end{equation}
The resulting amplitude reproduces well experimental  phase shifts and
inelasticities in the $I=J=0$ channel up to $1.2$ GeV.

\subsection{$\pi\pi$ in nuclear matter}

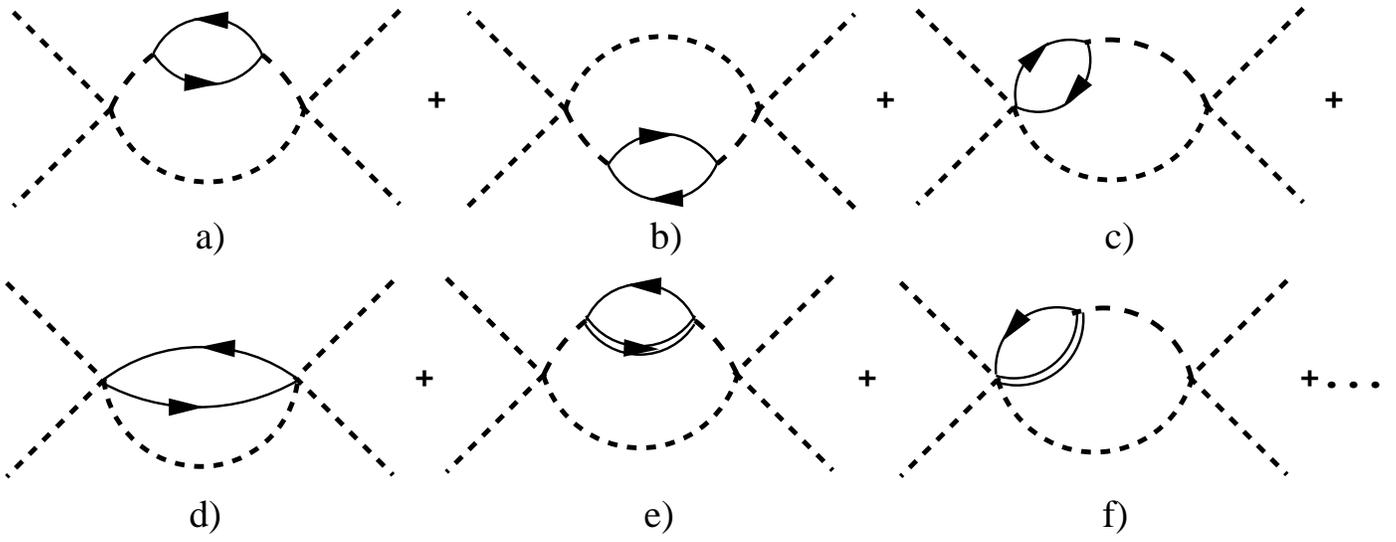
\begin{figure}
\caption{Some new terms of the Bethe-Salpeter equation in nuclear matter.
Bubbles represent particle-hole and $\Delta$-hole excitations.}
\label{fig:LSN} 
\end{figure}

In the nuclear medium, the pion propagators are strongly modified,
mainly due to their coupling to particle-hole $(ph)$ and $\Delta-$hole
$(\Delta h)$ excitations. Therefore, new terms, of the type depicted in
fig.\ \ref{fig:LSN} have to be included in the calculation of the 
scattering amplitude. It has been shown \cite{Cha:99,Chi:97} that the 
terms c),d),f),... of fig.\ \ref{fig:LSN} cancel the off shell
contribution of the  terms a),b),e),... so that only these
latter terms must be considered with the $\pi\pi$ amplitudes factorized
on shell. As a result the new BS equation is written as
\begin{equation}
\label{eq:LSN}
T_{\pi\pi} = V_{\pi\pi} + V_{\pi K} G_{KK} T_{K\pi} + V_{\pi\pi}
\tilde{G}_{\pi\pi}
T_{\pi\pi}.
\end{equation}
 As the $K\bar K$ state contributes very little to $T_{\pi\pi}$  at 
low energies, the vacuum values are kept for $G_{KK}$. Only $T_{\pi\pi}$ and 
$\tilde{G}_{\pi\pi}$ are renormalized in the medium. The function 
$\tilde{G}_{\pi\pi}$ is now given by 
\begin{equation}
\label{eq:PRP}
\tilde{G}_{\pi\pi} = i\int \frac{d^4 k}{(2 \pi)^4}
 \frac{1}{k^2 - m^2_{\pi} -\Pi(k)} 
\frac{1}{(P - k)^2 - m^2_{j}-\Pi(P-k)},
\end{equation}
where $\Pi(k)$ is the pion selfenergy in the nuclear medium calculated as
\begin{equation}
\Pi (k) = \frac{(\frac{f}{m_\pi})^2 \vec{k} \, ^{2}
U (k)  }{1 - (\frac{f}{m_\pi})^2 g' U (k) },
\end{equation}
with $g'$ the Landau-Migdal parameter ($g'=0.7$), $f$ the $NN\pi$ coupling
constant ($f=1$), and $U$ the Lindhard function for both $ph$ and $\Delta h$
excitations\cite{Ose:82}. In addition, we have also included the contribution
of $2p2h$ excitations as explained in ref. \cite{Chi:97}.

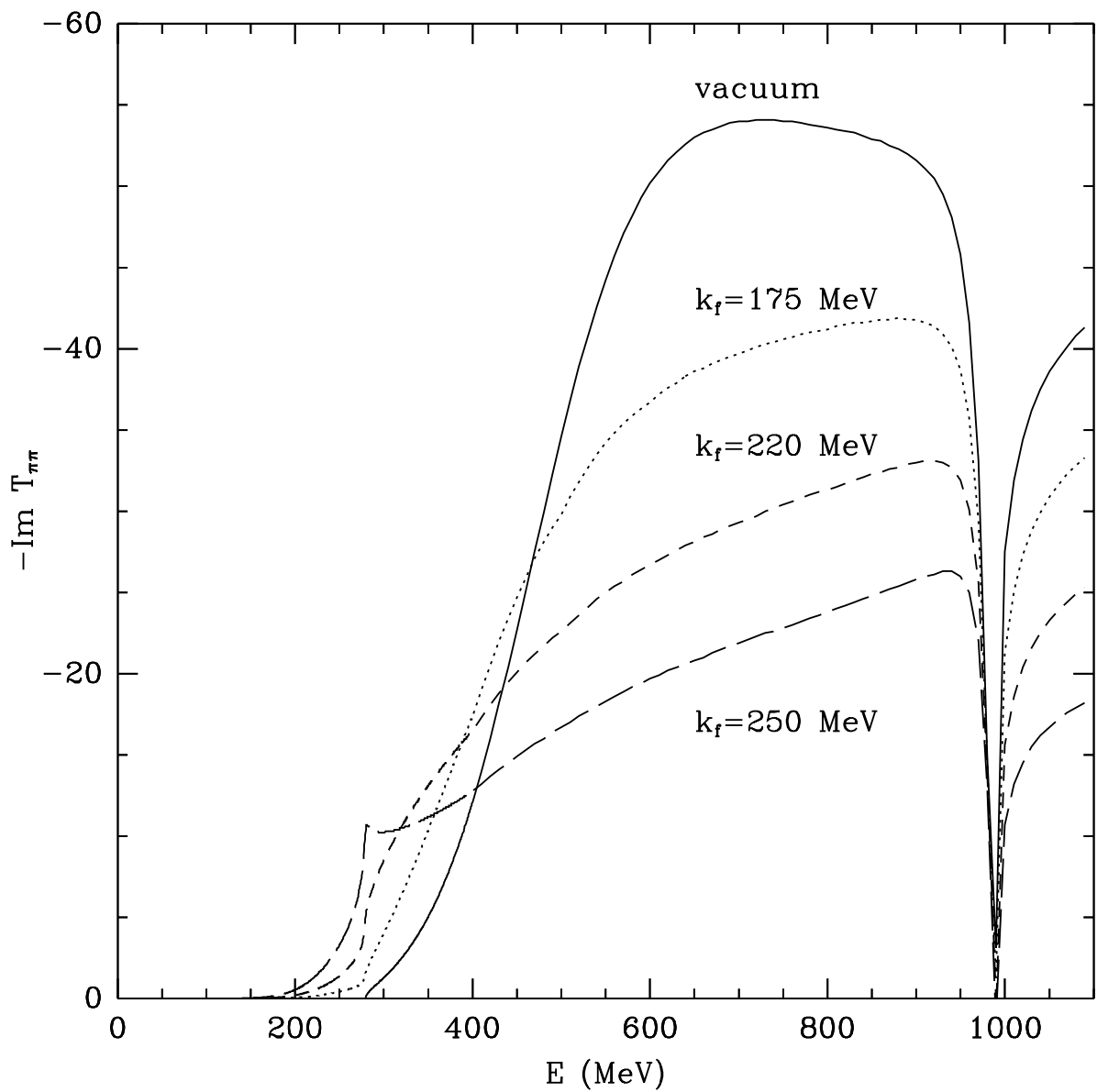
\begin{figure}
\caption{ Im $T_{\pi\pi}$ in the scalar isoscalar channel at different nuclear
densities as a function of the CM energy of the pion pairs. The labels
correspond to the nucleons Fermi momentum.}
\label{fig:IMT} 
\end{figure}

 In fig.\ \ref{fig:IMT} we show the imaginary part of the $\pi\pi$ scattering
amplitude for several nuclear densities. The results show a displacement of 
strength towards low energies. Although eq. \ref{eq:LSN} leads to a zero value 
for the amplitude close to the Adler's zero, ($s= m_\pi^2/2$ with $s$ the
Mandelstam variable of the two pions system), the accumulation of strength
close and below the two pion threshold is large and grows rapidly as a function
of the nuclear density.
 Very similar results were found in other works using different models for $\pi\pi$
interaction but imposing some minimal chiral constraints. See, for instance,
fig. 12 of  ref.  \cite{Aou:95} and fig. 3 of ref. \cite{Sch:98}.

\section{The $\pi N \rightarrow \pi \pi N$ reaction}
\label{sec:pin}

Our understanding of the $\pi N \rightarrow \pi \pi N$ reaction at low energies
has improved considerably in the last few years both experimentally and 
theoretically \cite{Ose:85,Jae:92,Ber:95,Jen:97,Bol:97}. The model we present
here follows closely that of  ref. \cite{Ose:85}, where only the  
$\pi^- p \to \pi^+ \pi^- n$ process was calculated. In this paper we will
extend it to other charge channels and will include  additional mechanisms,
that have been found to be important in the calculation of some differential
cross sections\cite{Sos:92,Bon:98}.

At low energies, the only particles that need to be considered are 
pions, nucleons, and the $\Delta$ and Roper resonances. The need of the latter
seems surprising, because of its relatively high mass  and the smallness of the
$\pi N N^*$ coupling. However, the Roper resonance partly decays  to
a nucleon and a s-wave pion pair. This mechanism does not vanish at threshold, 
unlike most other terms, and it has been found to be very important for
all charge channels that allow for an I=J=0 pion pair\cite{Sos:92}. 
 
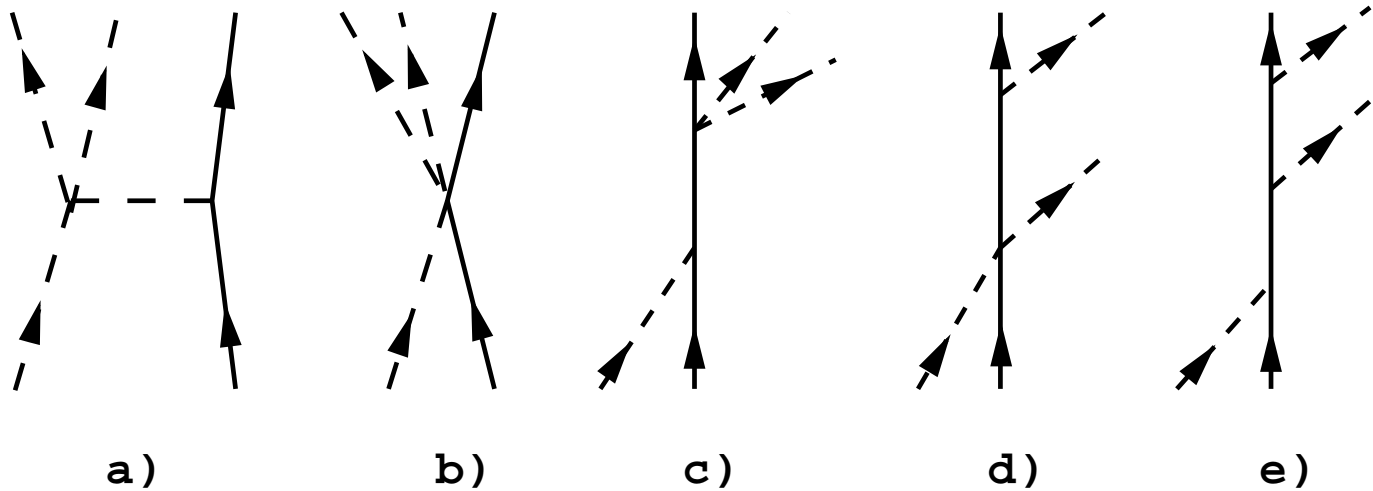
\begin{figure}
\caption{Feynman  diagrams contributing to the  $\pi N \rightarrow \pi \pi N$
reaction. Dashed lines are pions. External solid lines are nucleons. Internal
solid lines are nucleons,$\Delta$'s and $N^*(1440)$ where possible.}
\label{fig:DIA}
\end{figure}

We have considered in our model the mechanisms represented by the 
diagrams of fig. \ref{fig:DIA}. The dashed lines are pions, and the 
internal solid lines are all possible baryons ($N,N^*,\Delta$).
Notice, that different charge channels allow or forbid some of the
diagrams. Also all possible different time orderings, not explicitly 
depicted, are included in the actual calculation. The Feynman
rules used to calculate the scattering amplitude are derived from
the effective Lagrangians of the Appendix. It is relevant to mention here
that no attempt to fit the $\pi N \rightarrow \pi \pi N$ data has been made.
Rather, standard values for the coupling constants, obtained from different 
experiments and/or analysis, have been used.

\subsection{ Isospin amplitudes}
\label{sec:ISA}
The scattering amplitudes for the different $\pi N\to\pi\pi N$ charge channels
can be written as linear combinations of four isospin amplitudes 
$T_{2I,I_{\pi,\pi}}$. Here, $I$ and $I_{\pi,\pi}$ are the total isospin of the
system and the isospin of the two final pions. The expressions, for the 
channels we are interested in, are 
\begin{equation}
\label{eq:pm}
T(\pi^- p\to\pi^+\pi^-n)=-\frac{1}{\sqrt{45}}T_{32}+\frac{\sqrt{2}}{3}T_{10}
     +\frac{1}{3}(T_{11}-T_{31})\, ,
\end{equation}
\begin{equation}
\label{eq:00}
T(\pi^- p\to\pi^0\pi^0n)=\frac{4}{\sqrt{45}}T_{32}+\frac{\sqrt{2}}{3}T_{10}\, ,
\end{equation}
\begin{equation}
\label{eq:pp}
T(\pi^+ p\to\pi^+\pi^+n)=-\frac{2}{\sqrt{5}}T_{32}\, .
\end{equation}
Before proceeding, it is interesting to make some qualitative considerations.
The symmetry of the pion pair wave function implies that the orbital angular
momentum of the two final state pions will be even for the  $T_{10},T_{32}$ 
and odd for the $I_{\pi\pi}=1$ amplitudes. Thus, at low enough energies, 
where only  s-wave would dominate, both $T_{11}$ and $T_{31}$ should be 
negligible. On the other hand, the cross section for the 
$\pi^+ p\to\pi^+\pi^+n$ channel is much smaller than for the other two cases. 
Therefore  $T_{32}< T_{10}$, what implies that the amplitudes for the
$\pi^+\pi^-$ and for the  $\pi^0\pi^0$ case should be equal at low energies
and also that they are, essentially, dominated by the production of 
scalar isoscalar pion pairs.
As a consequence, the $\pi^0\pi^0$  cross section  should be about one half
the one for  $\pi^+\pi^-$, once the different thresholds are taken into account.
Of course, the energies at which there are experimental data are not that close
to threshold and one cannot neglect $T_{11}$ and $T_{31}$. In our calculation,
we will proceed as
follows. First, using the Lagrangians of the appendix we calculate the 
amplitudes for the different charge channels. Then, by means of eqs. 
\ref{eq:pm}, \ref{eq:00} and \ref{eq:pp}, we obtain $T_{10},T_{32}$, and the 
combination $T_{11}-T_{31}$, and finally, we modify $T_{10}$ by
including the final state interaction  of the $I=J=0$ pion pairs .

At low energies, due to the smallness of the $I_{\pi\pi}=1$ contributions, 
we only need to consider the effects of the FSI on the $I_{\pi\pi}=0,2$
channels. Furthermore, as it is shown in ref. \cite{Rap:99} the $\pi\pi$ 
interaction in for $I_{\pi\pi}=2$ is quite weak and does not change 
appreciably inside the nuclear medium. Thus only the FSI on the
$I_{\pi\pi}=0$ channel could produce large effects on the scattering
amplitude both in vacuum and in nuclei. To account for it, the amplitude 
$T_{10}$ is modified in the following manner,
\begin{equation}
\tilde{T}_{10}=T_{10}{\mathcal{F}}=T_{10}+T_{10} \tilde{G}_{\pi\pi} T_{\pi\pi}
\end{equation}
where $\tilde{G}_{\pi\pi}$ and $T_{\pi\pi}$ are the two pions propagator
and the scalar isoscalar two pions amplitude defined in eqs. \ref{eq:PRP}
and \ref{eq:LSN} respectively. Using again eqs. 
\ref{eq:pm}, \ref{eq:00} and \ref{eq:pp}, where $T_{10}$ is now replaced 
by $\tilde{T}_{10}$ we obtain the scattering amplitudes for the physical
channels.

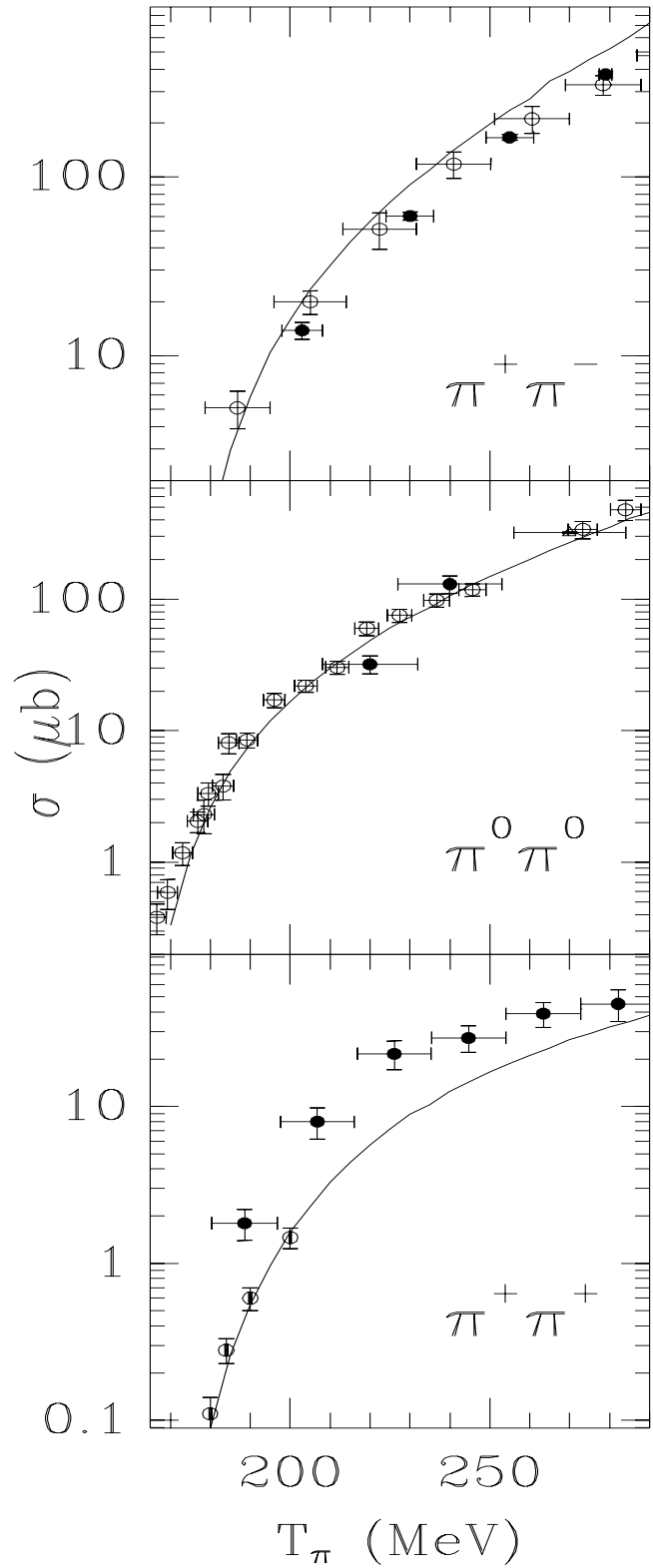
\begin{figure}
\caption{Total cross section for the $\pi N \to \pi \pi N$ reaction
vs. pion kinetic energy. Upper box: $\pi^- p \to \pi^+ \pi^- n$;
Experimental points from refs. \protect\cite{Ker:89,Bjo:80}. 
Middle box:$\pi^- p \to \pi^0 \pi^0 n$; Data from refs. 
\protect\cite{Low:91,Bel:80,Bun:77}. Lower box: $\pi^+ p \to \pi^+ \pi^+ n$; 
Data from refs. \protect\cite{Sev:91,Ker:90}.}
\label{fig:CRS}
\end{figure}

As can be seen in fig. \ref{fig:CRS}, this model, without free parameters,
reproduces fairly well the total cross sections for the three channels
considered. This agreement extends to differential cross sections, as it will 
be shown for some cases in next section. The net effect of the inclusion of
the pions FSI is a small enhancement of the total cross section
for the $\pi^+\pi^-$ and $\pi^0\pi^0$ channels, and slight
modifications of the differential cross sections, as it was expected,
given  that the $\mathcal{F}$ factor changes very smoothly, in vacuum, 
over the available phase space.

\section{The $\pi A \rightarrow \pi \pi X$ reaction}
\label{sec:nuc}

There are several nuclear effects that could modify he pion production cross 
section, like Fermi motion,
Pauli blocking, pion absorption and quasielastic scattering. Additionally,
there could be new reaction mechanisms involving more than one nucleon,
but the contribution of this kind of processes has been shown to be quite small
 \cite{Bon:98}. To account for the medium effects we follow ref.
\cite{Ose:86}. Assuming only one nucleon mechanisms, the cross section can be
written as
\begin{eqnarray}
\sigma=\frac{\pi}{q}\int d^2 \vec {b} \, dz \, F_{ISI}\,
       \int\frac{d^3\vec {k}}{(2\pi)^3} 
       \int\frac{d^3\vec {q_1}}{(2\pi)^3} 
       \int\frac{d^3\vec {q_2}}{(2\pi)^3} 
       n(|\vec {k}|) (1-n(|\vec{q}+\vec{k}-\vec{q_1}-\vec{q_2}|))\\
       \sum_{s_i s_f} |T|^2 \frac{1}{2 \omega(\vec{q_1})}
       \frac{1}{2 \omega(\vec{q_2})}
       \delta(q^0+E(\vec{k})-\omega(\vec{q_1})-\omega(\vec{q_2})-
                E(\vec{q}+\vec{k}-\vec{q_1}-\vec{q_2}))
       \, F_{out\_abs}\, ,
\end{eqnarray}
where $\vec{q},\vec{k},\vec{q_1}$, and $\vec{q_2}$ are the initial pion,
initial nucleon and two final pions momenta; $E(\vec{k}),\omega(\vec{q_1})$
and $\omega(\vec{q_2})$ are their energies. 
The spatial volume element is written in terms of $\vec{b}$, perpendicular, 
and $z$ parallel to the beam momentum.
The occupation number $n(.)$ refers to the local density, it takes the 
value 1 when the argument is below the local Fermi momentum 
($k_F=(3\pi^2\rho(\vec{b},z)/2)$) and zero above it. Note 
the sum over initial and final spins, but not over isospin because the
channels we study involve only protons in the initial state and neutrons in
the final one. In this paper, we only consider symmetric nuclei, and 
we will use the same density for neutrons and protons. The nuclear densities
are taken from ref. \cite{Jag:74}. The amplitude $T$ is also evaluated at the
local density, what fundamentally means that its only changes occur in the 
$\mathcal{F}$ factor, related to the $\pi\pi$ interaction. Finally,
the factors $F_{ISI}$, and $F_{out\_ abs}$ account for the flux loss 
due to pion absorption and scattering in the case of  $F_{ISI}$ and
for pion absorption alone in the case of $F_{out\_ abs}$.
 Both pion absorption and quasielastic scattering are quite 
strong at the energy of the incoming pion and both reduce the effective initial
pion flux. That is clear for the absorption, but it is also true for the  
$\pi$-nucleus quasielastic scattering, because in these collisions the pion
loses  always some energy, what reduces enormously the possibility of a
subsequent pion production. We implement the flux lost with the eikonal factor
\begin{equation}
F_{ISI}=\exp{ (
\int_{-\infty}^z dz'\,(P_{abs}(\omega(q),\rho)+P_{qua}(\omega(q),\rho)
        )},
\end{equation}
where $P_{abs}, P_{qua}$ are the absorption and quasielastic scattering 
probability per unit length, that we take from ref. \cite{Ose:86}.
Due to these interactions, the flux reaching the inner nucleus is quite small
and the reaction happens at the surface, what reduces considerably the 
possibility of  strong medium effects. 
As mentioned in ref. \cite{Rap:99}, the final pions have low energy and 
absorption or quasielastic scattering is a minor effect in their case. This 
has also been confirmed in the analysis of experimental data of ref. 
\cite{Cam:93}. Nonetheless, it could affect more those events happening at 
high densities in the center of the 
nucleus. Therefore, we include the factor $F_{out\_abs}$ that calculates
the reduction of the cross section due to the absorption of any of the
final pions. It is given by 
\begin{equation}
F_{out\_abs}=\exp{(
\int_{-\vec{b},z}^\infty dl_1\,P_{abs}(\omega(q_1),\rho)
       )}\, 
       \exp{(
\int_{-\vec{b},z}^\infty dl_2\,P_{abs}(\omega(q_2),\rho)
       )}\, ,
\end{equation}
where $dl_1,\, dl_2$ are the elements of longitude along the $\vec{q}_1$
and $\vec{q}_2$ directions. The probability of quasielastic collisions
is very small at these energies and we ignore it.

\section{Results}
The results that will be presented in this section are all compared with CHAOS 
data.  We have approximately implemented  the experimental acceptance
cuts in our calculation. Let us take the beam direction as the $z-$axis, and
$x$ forming with $z$ the horizontal plane, then
$\phi=0\pm 7$ degrees, where $\phi$ is the pion angle with the $xz$ plane, and
$10<\theta<170$, with $\theta$ the pion angle with the $z$ direction.
 Also, we have taken a pion kinetic energy threshold, $T_\pi > 11\, MeV$. 

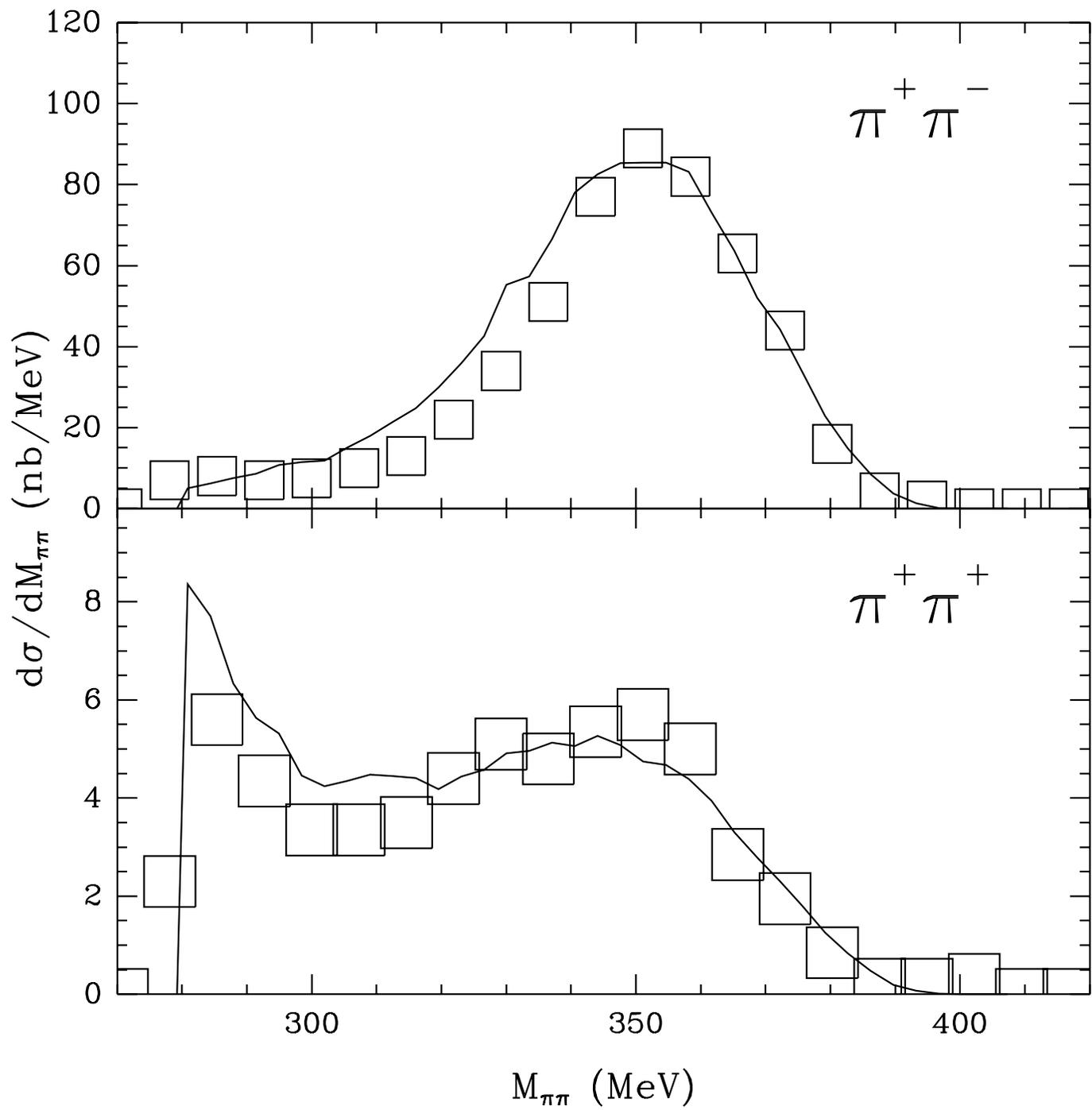
\begin{figure}
\caption{ Two pion invariant mass distributions in the 
$\pi^-+d\to \pi^+\pi^- nn$ (upper box), and $\pi^++d\to \pi^+\pi^+$ (lower box)
reactions. Experimental points are from  ref.   
\protect\cite{Bon:99}.}
\label{fig:DEU}
\end{figure}

In fig. \ref {fig:DEU},
we show the invariant mass distributions for the deuteron case. We obtain a
quite good agreement in the $\pi^+\pi^+$ channel in both shape and size.
The agreement is not as satisfactory for the $\pi^+\pi^-$ case. As can be seen 
in fig. \ref{fig:CRS}, the total cross section is slightly overestimated by our
model, and a similar situation  is found for the mass distribution. In order 
to compare easily with the experimental shape, we have renormalized our result 
reducing it by a 20\% and this is what is shown in the figure. Let us 
remember here, that no parameter has been 
adjusted, and some of them, like those related to the Roper properties are 
very uncertain. A quite small change of a few percent in the Roper coupling 
constants would produce a fine agreement in size.
It is interesting to understand the quite different behaviour of the two
channels under consideration. As it was already stressed in ref. 
\cite{Bon:98}, the  $\pi^+\pi^+$ case follows closely a pure phase space 
distribution. The two peaks of the figure respond only to the geometry of the
experimental apparatus, which favours clearly the situations in which either
the pions go together (low $M_{\pi\pi}$ ) or in opposite directions
(high  $M_{\pi\pi}$). The smallness of the $M_{\pi\pi}$ distribution in the
$\pi^+\pi^-$ reaction reflects the much richer structure of the amplitude,
and it is produced by the destructive interference of large pieces.
Consider, as an example, the mechanisms (I) $\pi^- p\to n^*\to n
(\pi\pi)_{I=J=0}$ and (II)  $\pi^- p\to n^*\to \Delta \pi \to n \pi^+\pi^-$.
The (I) amplitude has a constant sign, all over the available phase space. 
However, process (II) has an amplitude approximately proportional to the 
scalar product of the final pions momenta. Therefore, it changes sign
when passing from low to high invariant masses. In fact, not only is there
destructive interference at low masses, the constructive interference 
increases  the size of the high mass region.

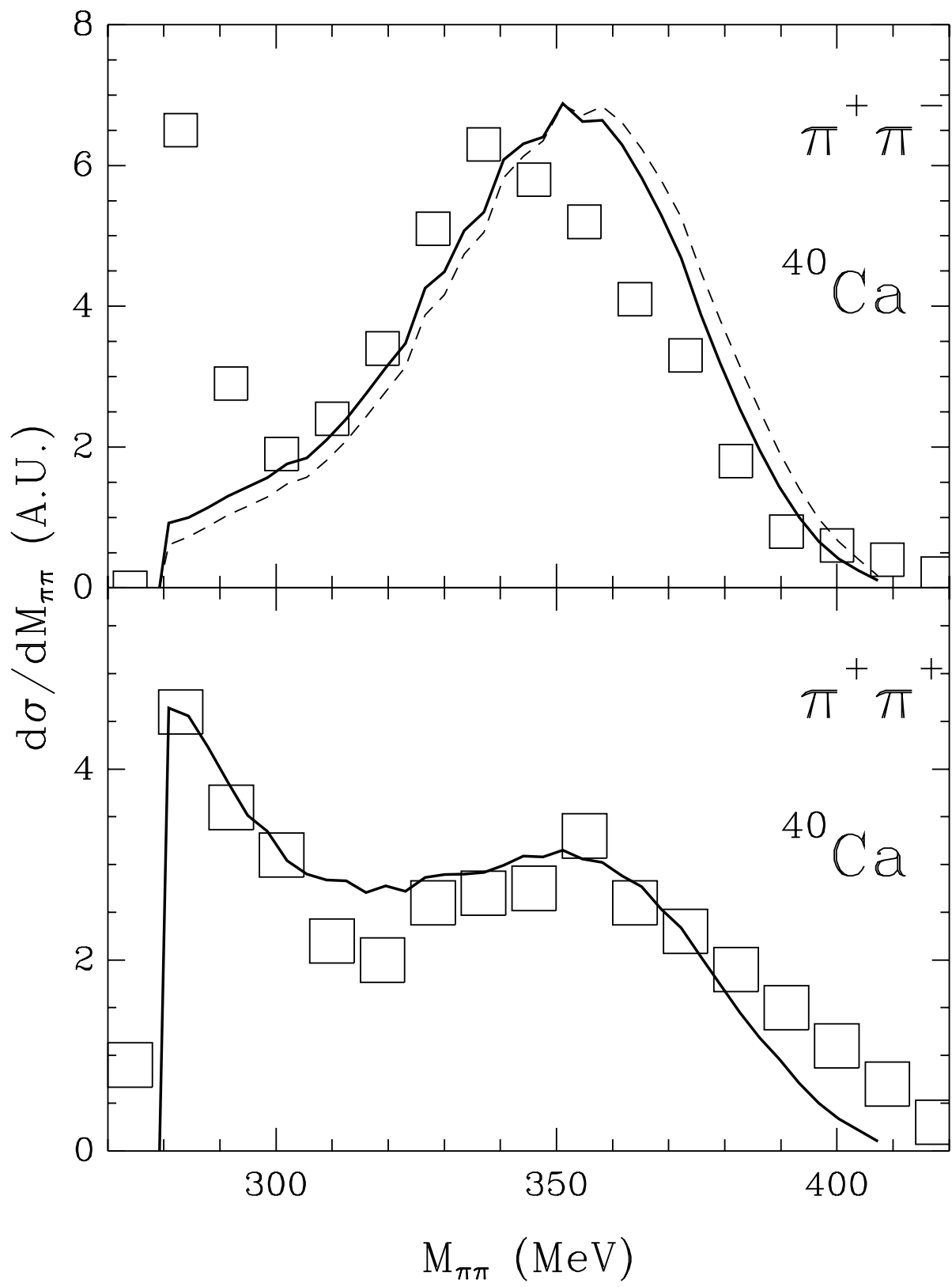
\begin{figure}
\caption{ Two pion invariant mass distributions in the 
$\pi^-+Ca\to \pi^+\pi^- X$ (upper box), and $\pi^++Ca\to \pi^+\pi^+ X$ (lower box)
reactions. Solid line, full calculation; dashed line, no medium effects in the
FSI of the two pions. Experimental points are from  ref.   
\protect\cite{Bon:96}.}
\label{fig:CAL}
\end{figure}

Our results for Calcium are shown in fig. \ref{fig:CAL}. Again, there is a very
good agreement with the $\pi^+\pi^+$ data. The shape of the figure is quite
similar to the deuteron case. There is some softening, and the distribution
reaches higher masses. Both features are mostly  due to the Fermi motion of
nucleons. Pion absorption is weak at low energies. Hence, it does not affect
much the final pions, although it is partly responsible for the reduction
of the spectral function  at high masses.
More important is the initial pion absorption, which apart from
changing the total cross section, prevents 
the pions from reaching the nucleus core, therefore decreasing the 
effective density at which the reaction happens, and modulating all nuclear
effects.

The quality of the agreement on the latter channel, gives us confidence
that all standard nuclear effects are properly taken into account. In particular
that we have a proper description of where in the nucleus the reaction takes
place. 

Finally, let us discuss the   $\pi^+\pi^-$ data, the channel in which 
the nuclear dependent $\pi\pi$ interaction in the scalar isoscalar channel
could be responsible of the presence of a low mass peak.
Our results show some enhancement close to threshold, even when  
the pion FSI is calculated in vacuum. This is due to Fermi motion. In the 
full calculation, the inclusion of the medium dependence in the FSI leads 
to a further enhancement and a displacement towards low masses, although is 
not enough to reproduce the data.

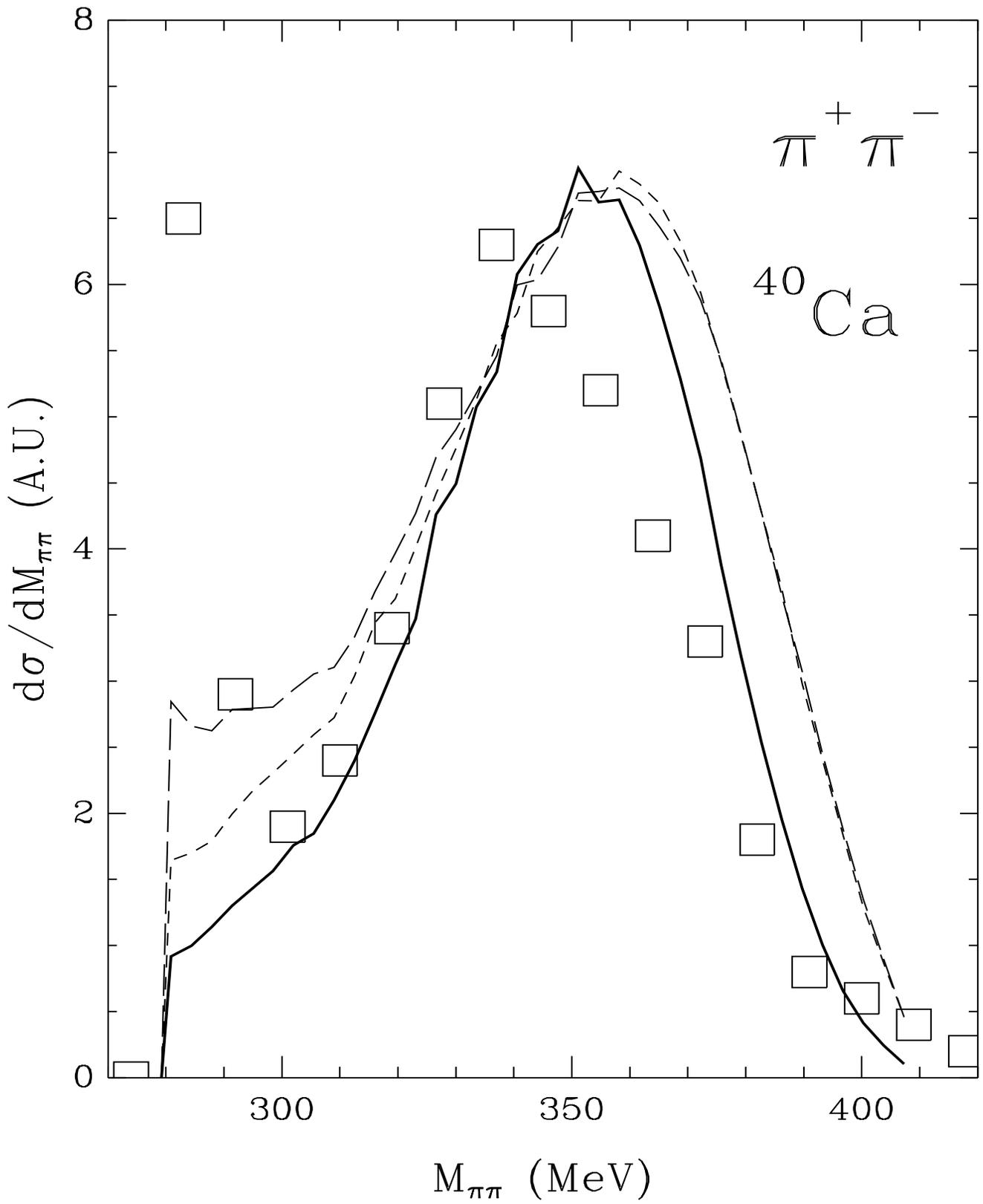
\begin{figure}
\caption{Same as fig. \protect\ref{fig:CAL}, using a fixed averaged density.
Short dashed line, $\rho=0.5\rho_0$; long dashed line  $\rho=0.7\rho_0$.
Solid line, full calculation}.
\label{fig:CAL2}
\end{figure}

One first question is the consistency  of our results with those of 
ref. \cite{Rap:99}, which reproduce well the experiment. In order
to do a proper comparison, we should impose some approximations that were used
there. In particular, we find that pion absorption forces the reaction to 
occur at the nuclear surface, at densities lower than those used in ref. 
\cite{Rap:99}. 
If we impose a high fixed average density (see fig. \ref{fig:CAL2}), 
we also get a large, although insufficient enhancement, and too high values 
for the mass distribution at high masses, produced by the excess of Fermi 
motion. 
If we neglect pion absorption, the results follow closely the 
curve corresponding to $\rho = 0.5 \rho_0$, as expected, given the fact that 
for the Calcium nucleus, the average density is approximately half the nuclear
density. 

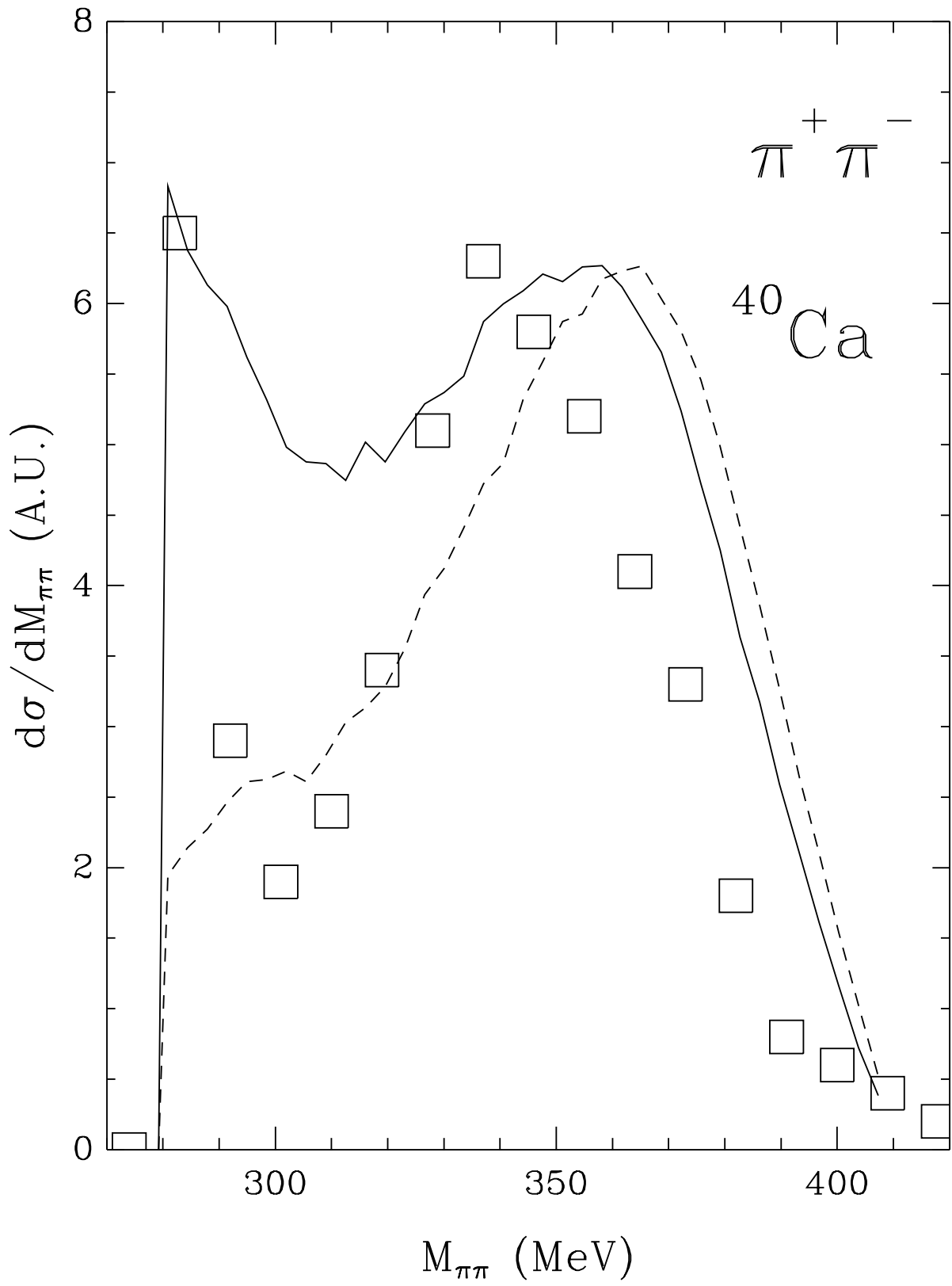
\begin{figure}
\caption{Same as fig. \protect\ref{fig:CAL}, using a fixed averaged density
 $\rho=0.7\rho_0$ and a simplified model (see text).
Dashed line, FSI in vacuum; Solid line, FSI in medium.}
\label{fig:CAL4}
\end{figure}

In order to facilitate further the comparison with the results of ref.
\cite{Rap:99} we have also considered a simplified model for the reaction,
neglecting all terms with a $\Delta$ resonance.
The two curves in fig. \ref{fig:CAL4} correspond to FSI calculated in vacuum
and with a density  $\rho = 0.5 \rho_0$.
 This simplified model has a similar structure to that of  ref. \cite{Rap:99}, 
and although it is able to reproduce the peaks structure in nuclei, it does so 
at the price of overestimating for the deuteron case the threshold region by a 
large factor.

\section{Conclusions}
We have studied the $A(\pi^+,\pi^+\pi^\pm)X$ reactions on the deuteron and on 
Calcium using a realistic model for the elementary $\pi N\to \pi\pi N$
production, and including several nuclear medium effects, like Fermi motion,
pion absorption, pion quasielastic scattering, and the medium dependent 
$\pi\pi$ interaction in the  scalar isoscalar channel. 

 We find a very good agreement with the experimental data  for the $\pi^+\pi^+$
production, which gives us confidence on our treatment of the common nuclear
effects. However, we are unable to reproduce fully the strong enhancement 
close to the two pion threshold found in the experiment for the 
$\pi^+\pi^-$ production. We also find that approximations previously used in 
the literature could be critical in reproducing such an enhancement. 
Several possibilities open up. It could happen that the $\pi\pi$ interaction in 
the $\sigma$ channel has a  stronger dependence on density than that provided 
by existing models. This would be most interesting. However, we have also found that
the smallness of the spectral function in the deuteron, close to threshold, 
is due to destructive interference between large pieces of the amplitude.
If some of these pieces is substantially modified in the nuclear medium,
the interference could disappear, and the spectral function would look 
more like phase space, as it is the case for $\pi^+\pi^+$ production.
This would be enough to reproduce the experimental data.
New dedicated experiments, with a wider phase space, and at lower energies,
where the interference effects are smaller are important to settle these
questions.

 We also find  that absorption of the incoming pion, leads the reaction to occur
at the surface, therefore reducing the signals of any density dependent effect.
Electromagnetically induced reactions, free from such a disadvantage are
clearly called for to investigate the medium effects on the $\pi\pi$ interaction
in the $\sigma$ channel

\acknowledgments
One of us (M.J.V.V.) would like to acknowledge useful discussions with
L.L. Alvarez-Ruso, N. Grion and P. Camerini. This work has been partially 
supported by DGYCIT contract no. PB-96-0753.

\appendix

\section{Lagrangians used in the $\pi N \rightarrow \pi \pi N$ model}

\subsection{Pions and nucleons}

 The scattering amplitude of the mechanisms depicted in fig. \ref{fig:DIA}, 
considering only nucleons for the internal lines, can be derived 
from the chiral lagrangians with the inclusion of baryons of ref. 
\cite{Gas:85,Mei:93,Pic:95,Eck:95}. 
The relevant pieces can be written as
\begin{equation}
L=L_{\pi\pi}+L_{\pi N}\, ,
\end{equation}
In this equation, $L_{\pi\pi}$ contains the purely mesonic interaction and $L_{\pi N}$
the meson(s)-nucleon terms. The mesonic part is given by
\begin{equation}
L_{\pi\pi}=\frac{f^2}{4} <\partial_\mu U^\dagger\partial^\mu U
+\chi (U+U^\dagger)> ,
\end{equation}
The brackets indicate the sum in flavour space. The matrices $U$ and $u$ 
are defined by
\begin{equation}
U(\Phi)=u(\Phi)^2=\exp\left\{ i \sqrt{2} \Phi / f  \right\}
\end{equation}
where 
\begin{equation}
\begin{array}{l}
\Phi \equiv \frac{1}{\sqrt{2}} {\btau}{\bphi}= \left(
\begin{array}{cc}
\frac{1}{\sqrt{2}} \pi^0  &   \pi^+  \\
\pi^-                     & - \frac{1}{\sqrt{2}} \pi^0 \\
\end{array} \right) \; , 
\end{array}
\end{equation}
where ${\btau}$ are the Pauli matrices and the $\pi^.$ are the pion 
fields. At lowest order, $f$ is equal to the pion decay constant,
$f=f_\pi=92.4 MeV$. Assuming isospin symmetry, the mesonic mass matrix is
\begin{equation}
\chi = \left( \begin{array}{cc}
m_\pi^2 & 0 \\
0 & m_\pi^2 \\
\end{array} \right) \; ,
\end{equation}

The interaction with the nucleon is described by the term
\begin{equation}
L_{\pi N}=\bar{\Psi}(i \gamma^\mu \nabla_\mu -M +\frac{g_A}{2}\gamma^\mu 
\gamma_5 u_\mu)\Psi .
\end{equation}
Here, $M$ is the nucleon mass, $\Psi$ is the nucleon isospinor 
\begin{equation}
\Psi = \left( \begin{array}{c}
  p\\
  n \end{array}  \right)\, ,
\end{equation}
and $g_A\approx 1.26$  is related to the pion nucleon coupling constant
($f_{NN\pi}/m_\pi$) by the Goldberger-Treiman relation
$\frac{f_{N N \pi}}{m_{\pi}} = \frac{g_A }{2 f}\,  $.
The covariant derivative of the nucleon field $\nabla_\mu \Psi$  is given by
\begin{equation}
\nabla_\mu \Psi = \partial_\mu B + \Gamma_\mu B \, , \quad
\Gamma_\mu = \frac{1}{2} \left (u^\dagger \partial_\mu u + u \partial_\mu 
u^\dagger \right) \, .
\end{equation}

Expanding $U$ and $u$, and keeping the terms with up to four pion 
fields, the following set of Lagrangians are obtained,
\begin{eqnarray}
\label{eq:pipi}
&{L}_{\pi \pi \pi \pi}& = \frac{1}{6f^{2}_{\pi}} \left[
(\partial_{\mu} \bphi \bphi)^{2}-\bphi^{2} 
(\partial_{\mu} \bphi)^{2} +\frac{1}{4} m_{\pi}^{2} \bphi^{4} \right] \,,
\\[0.4cm]
\label{eq:nnpi}
&{L}_{NN \pi}& =-\frac{f_{N N \pi}}{m_{\pi}} 
\bar{\Psi}\gamma^{\mu}\gamma_{5} \partial_{\mu} \bphi {\btau} \Psi \,,
\\[0.4cm]
\label{eq:nnpipipi}
&{ L}_{NN \pi \pi \pi}& = \frac{1}{6f^{2}_{\pi}}
\frac{f_{N N \pi}}{m_{\pi}} 
\bar{\Psi} \gamma^{\mu} \gamma_{5} [(\partial_{\mu} \bphi \btau) \bphi^{2}
- (\bphi \btau) (\partial_{\mu} \bphi \bphi)] \Psi\,,
\\[0.4cm]
\label{eq:nnpipi}
&{ L}_{NN \pi \pi}& = -\frac{1}{4 f^{2}_{\pi}} \bar{\Psi} \gamma^{\mu}
\btau (\bphi {\scriptstyle{\times}} \partial_{\mu} \bphi) \Psi \,.
\end{eqnarray}

Whereas the expression of the  ${L}_{NN \pi}$ Lagrangian is quite standard,
different forms can be found in the literature for the\  \ref{eq:pipi}
and\ \ref{eq:nnpipipi} pieces, because different representations for the 
pion field $U(\Phi)$ had been used. However, it can be shown that they produce 
the same physical amplitudes. In particular, they are equivalent to 
the Weinberg Lagrangians used in ref. \cite{Ose:85} when the chiral symmetry
breaking parameter $\xi$ is taken to be zero.

The ${ L}_{NN \pi \pi}$ term describes the isovector part of the 
$\pi N \to \pi N$ s-wave interaction. The very small isoscalar part
would appear at higher orders of the chiral Lagrangian. 
Adding a phenomenological isoscalar part, and doing a low energy approximation,
the Lagrangian  can be recast in the typical form 
\begin{equation}
L_{NN \pi \pi} = -4\pi \left\{ \frac{\lambda_1}{m_{\pi}}\bar{\Psi}\bphi^2 \Psi +
\frac{\lambda_2}{m_{\pi}^2} \bar{\Psi}
\btau (\bphi {\scriptstyle{\times}} \partial_{t} \bphi) \Psi \right\} \,.
\end{equation}
where  $\lambda _2= \frac{m_{\pi}^2}{16 \pi f^2_{\pi}}\approx 0.045$.
Fitting the constants to the  s-wave $\pi N$ scattering lengths
\cite{Koc:86}, one gets an slightly larger $\lambda _2=0.52$, 
and $\lambda_1 = 0.048$.

Finally, the pion and nucleon propagators are 

\begin{equation}
D_\pi (q) = \frac{1}{q^2 - m_\pi^2} \,,
\end{equation}
\begin{equation}
G_N (q) =  \frac{M}{E_q}\frac{1}{q^0-E_q+i\epsilon} \, ,
\end{equation}
where $E_q=\sqrt{q^2+M^2}$ 

\subsection{$\Delta$ resonance}
Diagrams d) and e) of fig. \ref{fig:DIA} could also have $\Delta$ resonances
as intermediate steps. The required $\Delta N\pi$ and $\Delta\Delta\pi$ vertices
are described by the following phenomenological Lagrangians
\begin{equation}
L_{\Delta N \pi} = \frac{f^*}{m_\pi} \psi^{\dagger}_\Delta 
S^{\dagger}_i (\partial_i\bphi) {\mathbf{T}^{\dagger} \psi_N }\, + \, h. c. \,,
\end{equation}
\begin{equation}
L_{\Delta \Delta \pi} = \frac{f_\Delta}{m_\pi} 
\psi^{\dagger}_\Delta S_{\Delta i} (\partial_i\bphi) 
\mathbf{T}_{\Delta} \psi_{\Delta}\,,
\end{equation} 
where the $\psi^{\dagger}_{\Delta (N)}$ are two-component spinor fields,
$\mathbf{S^{\dagger}}$ ($\mathbf{T^{\dagger}}$), and $S_{\Delta}(T_\Delta)$
are the spin (isospin) $1/2 \to 3/2$  and  $3/2 \to 3/2$ transition operators.  
Definition, and some useful algebraic relations of this operators can be
found in ref. \cite{Alv:98}.

The $\Delta$ propagator is given by
\begin{equation}
G_\Delta (p)=\frac{1}{W- M_\Delta + \frac{1}{2} i \Gamma_\Delta (p)} 
\end{equation} 
where $W$ is the $\Delta$ invariant mass and the resonance width 
$\Gamma_\Delta$ is
\begin{equation}
\label{eq:Dwi}
\Gamma_\Delta (W)= {\frac{1}{6\pi}} \left ( {\frac {f^*}{m_{\pi}}} \right )^2
        {\frac{M}{W}}\, |{\mathbf{q_{cm}}}|^3 \,\Theta(W-M-m_{\pi})\,.
\end{equation}
with $\mathbf{q_{cm}}$  the pion momentum in the resonance rest frame.
The $\Delta N\pi$ coupling constant, obtained from $\pi N$ phase shifts or from 
the experimental width, using eq. \ref{eq:Dwi}, takes the value $f^*=2.13$.
For the $\Delta \Delta \pi$ coupling we take the quark model value, 
$f_\Delta=4/5 f_{NN\pi}$ \cite{Bro:75}.

\subsection{Roper resonance}

The Roper resonance can decay into a nucleon and a pion, a  $\Delta$ and a pion
and a nucleon and two s-wave pion. The effective Lagrangians we use to describe
these interactions are
\begin{equation}
\label{eq:NNp}
 L_{N^* N \pi} = -\frac{\tilde{f}}{m_{\pi}} 
\bar{\Psi}_{N^*} \gamma^{\mu}\gamma_{5} \partial_{\mu} \bphi \btau \Psi_N \, 
+ \, h. c. \,, 
\end{equation}
\begin{equation}
L_{N^* \Delta \pi} = \frac{f_{N^* \Delta \pi}}{m_\pi}
\psi^{\dagger}_{\Delta} S^{\dagger}_i (\partial_i \bphi) {\mathbf{T}^{\dagger}} 
\psi_{N^*} \, + \, h. c. \,.
\end{equation}
and
\begin{equation}
 L_{N^* N \pi \pi} = - C
\bar{\Psi}_{N^*} \bphi \bphi \Psi_N \, + \, h. c. \,.
\end{equation}

The coupling constants $\tilde{f},f_{N^* \Delta \pi}$ and $C$ are calculated
from the $N^*$ width and branching ratios. There are considerable uncertainties
in the experimental information, but using the central values from ref. 
\cite{Cas:98} one gets $\tilde{f}=0.477, f_{N^* \Delta \pi} = 2.07 $ and 
$C=2.3 m_\pi ^{-1}$ \cite{Alv:98}.
The Roper propagator is given by
\begin{equation}
G_{N^*} (p)=\frac{1}{W- M^* + \frac{1}{2} i \Gamma_{N^*} (p)}  
\end{equation}
At low energy, the width is dominated by the decay into the $N\pi$ channel,
and it is given by
\begin{equation}
\Gamma_{N^*}\approx \Gamma_{N\pi}={\frac{3}{2\pi}} \left 
( {\frac {\tilde{f}}{m_{\pi}}} \right )^2
{\frac{M}{W}}\, |{\mathbf{q_{cm}}}|^3 \,\Theta(W-M-m_{\pi})\, .
\end{equation}

Whereas all other constants are obtained from different experiments, without
two s-wave pions in the final state, and hence are independently
determined, we should handle with more care $f_{N^* \Delta \pi}$ and $C$.
Both constants correspond to processes with two pions in the 
final state, which could be in the scalar isoscalar channel. 
The constants have been fitted to the experimental Roper partial widths
using the Born approximation (without final state interaction of the pions),
and therefore effectively incorporate already the vacuum renormalization due
to the final state interaction of the pions. As we will include the FSI
explicitly in our calculation, we have first to discount its effects on these
two constants to avoid double counting.

 As explained in Sec. \ref{sec:ISA} our model implements FSI by multiplying 
the scalar isoscalar part of the amplitude by the factor ${\mathcal F}$. 
This factor depends smoothly in vacuum on the invariant mass of the two pion 
system, and is practically constant over the available phase space region
of the $N^*(1440)$ decay. We thus take a corrected value for $C$
\begin{equation}
C\to \frac{C}{|\bar{\mathcal F}|},
\end{equation} 
where  $\bar{\mathcal F}$ is the value of ${\mathcal F}$ at the average 
invariant mass of the two pions in the  $N^*(1440)$ decay. 
 The correction is not as important for the $f_{N^*\to \Delta \pi}$ case.
First, because only a small part of the partial width produces 
pion pairs in the scalar isoscalar channel, and second 
because at the low energies of the experiment under analysis, the contribution 
of this  kind of mechanisms is small.

\end{document}